\documentclass[a4paper,12pt,twocolumn]{article}

\usepackage{graphicx}
\usepackage{txfonts}
\usepackage{longtable}
\usepackage[font=small,labelfont=bf]{caption}
\usepackage{abstract}
\usepackage{fullpage}
\def \etal {\textit{et al. }}
\def \apj {ApJ }

\def \mnras {MNRAS }

\begin{document}
\title{KeSeF - Kepler Self Follow-up Mission\\
\small{White paper on an alternate science investigations for the Kepler spacecraft}}

\author{Aviv Ofir \\ \small{Institut} f\"ur Astrophysik, Georg-August-Universit\"at, G\"ottingen, Germany. \\ \small{avivofir@astro.physik.uni-goettingen.de}}

\date{}

\twocolumn[
\maketitle

\begin{onecolabstract}
The Kepler spacecraft is currently unable to hold a steady pointing and it is slowly drifting during observations. We believe that if one has to deal with targets that drift across the CCDs, one should at least be able to track the targets well enough to correct for some -- if not most --  of the problems caused by this drift. We therefore propose to observe as many stars as possible in short cadance. We propose that at least all currently known planetary candidate host stars will be so observed, with possibly known Kepler eclipsing binaries, astroseismology targets, guest observer targets and new targets in increasingly lower priority. We also outline the modifications needed to flight software in order allow for such observations to take place, aiming to provide ample non-photometric data that should allow post-processing to recover most of the pre-failure photometric performance. In total, the KeSeF Mission will allow Kepler to follow up it's own previous discoveries in a way that is not otherwise possible. By doing so it will enable to continue pursuing nearly all the science goals that made the original mission choose staring at a single field of view in the first place.

\end{onecolabstract}
]

\section{Introduction}

The failure of two of Kepler's four reaction wheels made the spacecraft unsuitable for the survey which it was designed to perform. Additionally, some of Kepler most important discoveries are the large number of candidates, including the large number of multi-transiting systems. Continued monitoring of these systems is scientifically very desirable, allowing the detection of long-period planets as well as an increased time baseline for transit timing variation (TTVs) signals among other things (see details on \S \ref{Rational}). Continued observations of all of Kepler's $~150,000$ targets, even at the reduced precision possible with the crippled spacecraft, is unfeasible. However, it appears possible (see \S \ref{Implementing}) that observing a relatively small subset of the stars -- primarily the already-identified Kepler Objects of Interest (KOIs) -- at precision not dramatically different than before will be possible. Here we propose the Kepler Self Follow-up  (KeSeF) Mission to use Kepler itself for its own follow up, aiming to maximize the scientific return \footnote{Literaly, "kesef" means "money" in Hebrew, again stressing the maximizing return-on-investment theme.} on the original Kepler Survey Mission.


\section{Science Rational}
\label{Rational}

\textbf{Target stars:} It is very fortunate that the number of foreseen targets for the KeSeF mission (minimum of 5000 targets, see \S \ref{Implementing}) is larger than the current number of KOI (Kepler Object of Interest) host stars, and having more KeSeF targets may allow also the inclusion of all known Kepler eclipsing binaries (that may also include circumbinary planets). We therefore label the two populations above as primary and secondary target populations, respectively, that the proposed KeSeF mission aims to observe and perform the self follow-up on. Remaining target resources, if available, may be given to other investigation such as astroseismology, guest observer programs, high-quality non-candidate stars, etc. A small number (few per Kepler channel) of hand-picked particularly bright and stable targets may be added as a calibration set.

\textbf{Extended baseline:} Continuing the Kepler Survey Mission with KeSeF will allow to explore the thousands of candidate planetary systems to the fullest extent possible. It is difficult to give exact science deliverables from the proposed KeSeF Mission as the achievable photometric precision is yet unknown. Still, all the basic reasons that caused the Kepler Extended Mission to be a continued monitoring of the same Field of View (FOV) are still valid here, among them:
\begin{itemize}
\item long-period planets, primarily giant planets but perhaps smaller too (depending on available precision)
\item Smaller planets may become detectable with the addition of new data points.
\item Increased time baseline for transit timing variations (TTVs). We note that the contribution of extending the baseline to TTVs signals is much more than linear ($\propto t^{5/2}$ when uniformly sampled), making the continued monitoring of the same FOV particularly valuable. 

A good example is KOI 1574 (see Fig. \ref{KOI1574}): this system contain a giant inner planet in a 114d period and an interesting outer super-Earth-mass and low-density planet in a 191d orbit where both masses, and hence densities, were derived from TTVs. Importantly, the system includes one of the largest TTV signals known (amplitude of $\sim8$ hr) but the long periods, coupled with the inactive module on Kepler, caused the 191d planet to be observed in transit only 4 times during the entire mission (no more events for this planet beyond those shown). Such a small number of events make dynamical model fitting prone to over fitting and systematic errors as the number of free parameters is similar to the number of data points - so every new data point makes a very significant contribution to long-period planet studies. Also, one can see that the model points for the outer planet (red $\triangle$ on Fig. \ref{KOI1574}) actually have a different mean period than the observed one (manifested the upwards trend in (O-C) figure) - again showing the limitation of small number of data points.

\begin{figure}[htb]\includegraphics[width=0.5\textwidth]{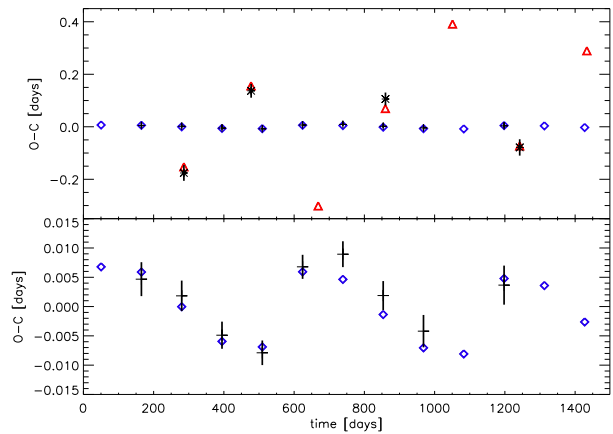}
\caption{Top: Observed transit timing variations of two planets around KOI 1574 through quarter 13 (black '$+$' symbols for TTV of the inner planet due to perturbations of the outer, black '$\ast$' symbols for TTV of the outer planet due to perturbations of the inner planet) compared to the best fit simulated ones ($\Diamond$ and $\triangle$ respectively). Error bars are indicated with thick lines. Bottom: Zoom-in showing the low amplitude TTVs of the inner planet (figure extracted from [6]).} 
\label{KOI1574}
\end{figure}

\item Also related to the above - the continued monitoring of eclipsing binaries (EBs) may allow the detection of non-transiting giant planets via binary eclipse time variations. Again, extending the baseline to such timing signals has a more than linear effect on the sensitivity to such signals.
\item Extending the time baseline for EBs may allow also the detection of new transiting circumbinary planets, as this already well-aligned population of targets is a particularly rich sample of targets [2]. Triply-eclipsing systems, such as KOI-126 [3] will also become more detectable.
\end{itemize}

Thus the proposed KeSeF mission has a high potential to continue and extend the great achievements from the Kepler Survey Mission.

\section{Implementing KeSeF}
\label{Implementing}

\subsection{General requirements}

The KeSeF proposal in unlike regular proposals in that it must be executed by a crippled spacecraft. Importantly, currently there are only a few and blurry details available about Kepler's expected photometric performance (e.g. a factor $>30$ in predicted performance for the same target, depending on yet unknown behavior [1]). We therefore turn to general past experience in astronomy for guidance: if high raw precision is available, and enough external data in given in order to attenuate external effects significantly, a large fraction of the original raw precision can be later recovered in post-processing. We therefore aim to minimize the effects of Kepler's coarse pointing ability while maximize the power of post processing. Furthermore, we try to limit ourselves only by hard limits that cannot be changed (e.g., volume of downloaded data) and not by softer limits like software-imposed limits.

The key enabling idea is that if one has to deal with targets that drift across the CCDs, one should at least be able to track the targets well enough to correct for some -- if not most --  of the problems caused by this drift. Such tracking can be achieved by observing only a small subset of the targets, but observing all of them at short cadance (1 minute). Our proposal is therefore made out of two main modifications to Kepler's flight software:

\begin{enumerate}
\item Adding the ability to specify targets (and apertures) in celestial coordinates so that the on-board software will do the coordinates-to-pixels translation on the fly and before each and every exposure. This will allow the number of pixels allotted to a given target to be maintained relative to the corresponding number during the Kepler Survey Mission.
\item The current limit of 512 simultaneous SC (short cadance) targets would be lifted, or at least significantly increased to $\sim 5000$ (required) or $\sim 10000$ (goal) simultaneous SC targets.
\end{enumerate}

Importantly, significant testing of the expected results of KeSeF can be made by implementing target tracking only, which we believe to be the easier of the two proposed modifications, on the 512 available SC targets. The light curves produced in this manner can give very good idea as for the expected performance of the full KeSeF Mission.

\subsection{Some details}

We note that some details can be given on how (and why) KeSeF may be implemented:

\textbf{Memory, computing and telemetry budgets:} The factor of x30 in the number of pixels read by changing a LC (long cadance) target to SC must by compensated by a corresponding reduction in the number of targets to keep the computational/telemetry load roughly as before, hence we require that at least 5000 LC targets will be available. Furthermore, one may be able to use the fact that the pixel data is now much more self-similar (more pixels of the same targets) to achieve better compression, enabling to transmit a larger number of SC targets with the current telemetry envelope, hence the 10000 SC targets goal. Also, by observing all the current KOIs the typical proposed target will be similar to the current typical target, hense the total pixel budget will not be significantly changed.

The current photometric apertures already include the required flexibility in post-processing since all targets already have one or two pixel halos beyond the optimal aperture. In KeSeF operation the used pixels for the apertures will always be no more than 1 pixel away from optimal ones (since they are tracked) and the target apertures will therefore need not be changed.

\textbf{Mission duration:} The observing mode will be as similar to the Kepler Survey Mission as possible: a fixed FOV (field of view) up to the final drift limit, followed by a momentum reset and re-pointing to the same starting point as before -- for as long as possible. Doing this may also be very repeatable and stable in some sense once perfected.

\textbf{Enabling significant post processing:} The pixel crossing time will be roughly 5 minutes, while transits are typically much longer (several hours). The former means that there will be several data samples during the crossing of even a single pixel - and these very pixels will be crossed again and again every nodding (=momentum reset) cycle, helping post-processing. The significantly different time scale mean that correcting the drift effects will likely have a small residual impact on transit detection.

\textbf{Post processing:} It is impossible, at this stage, to give details of the best post-processing technique. We do note that some techniques (e.g., SARS [4], TFA-EPD [5]) already allow the inclusion of external information, such as the pixel- and sub-pixel positions, in the decorrelation process. We therefore believe that an effective post-processing technique can be quickly identified.

\textbf{Other notes:} On-board targets tracking and pixel allocation must be able to account for sources that go on- and off- any of the CCDs. If a nodding limit will be greater than the size of the CCD is chosen - all targets would fall off- and on- the CCDs (and maybe on- or off- the next CCD in the same nodding cycle).

\subsection{Additional goals}
Below we give some incentives to explore the full range of possible implementation scenarios for the KeSeF mission, beyond the basic requirements:

\begin{itemize}
\item Having more target resources will allow inclusion and follow-up on all Kepler objects that were referred to in scientific publication (incl. pulsating stars , non stellar objects and continued guest observer program).

\item Even shorter integration times (shorter than 1 minute) should be explored. This will limit the number of targets, but may allow for better photometric performance.

\item One may allow including targets that were not previously observed (e.g., those that happen to fall outside or in gaps between CCDs on the nominal FOV). This will be possible since some of the nominal target will be falling off the CCDs in the same time.

\item We foresee that all targets for the KeSeF Mission would be in SC, but some users may opt to use LC data in order to maximize their number of targets - so maybe LC targets may not be completely deprecated.

\item Since the mode of operation we propose for KeSeF is very similar to the nominal one, we propose to keep many of the previous facilities as well, such as: guest observer programs, continuous pixel table up keeping, etc.

\end{itemize}

\section{Conclusions}
\label{Conclusions}

Above we presented an alternate science investigation for the Kepler spacecraft - follow-up observations of at least all the KOIs, and possibly many more interesting targets, that were detected during the Kepler Survey Mission. We described an operational mode that may allow regaining much of Kepler's lost photometric precision, allowing the follow-up observation proposed to be truly unobtainable in any other way. Such observations can contribute significantly to the understanding of these systems and to the science result of Kepler in General. We gave a real example that very closely resembles the most optimistic case desired: TTVs of a large inner planet allow to deduce the mass of an outer and small planet near- or in- the habitable zone. It is easy to see in this example how continued monitoring would improve the understanding of such high-value systems. This example is also from a rather faint target (KepMag=14.6) - one which would probably not be selected if the number of SC targets would remain 512, or $\sim10\%$ of the \textit{minimal} number we propose here.

Kepler was an outstanding exoplanets mission because it was carefully designed to do just one thing very well. We proposed the KeSeF Mission that will keep Kepler doing this exact thing and in a manner as similar as possible to the original mission: same field of view, same targets, same photometric apertures, same volume of data, same integration times, some environment (e.g. Solar radiation pressure) and so on. KeSeF does require some adaptations to flight software, but we believe these are both doable and worthwhile. Doing so will keep Kepler closely aligned with the original Kepler Mission goals, as well as with the Kepler Extended Mission goals, in the most natural way.

\section{Acknowledgments}
I would like to thank Eric Ford for reading and improving this proposal.

\section*{References}
\flushleft
\scriptsize{

[1] Call for White Papers for Alternate Science Investigations for the Kepler Spacecraft.

[2] Welsh, W.~F., Orosz, J.~A., Carter, J.~A., et al.\ 2012, Nature, 481, 475 

[3] Carter, J.~A., Fabrycky, D.~C., Ragozzine, D., et al.\ 2011, Science, 331, 562

[4] Ofir, A., Alonso, R., Bonomo, A.~S., et al.\ 2010, \mnras, 404, L99 

[5] Bakos, G.~{\'A}., Torres, G., P{\'a}l, A., et al.\ 2010, \apj, 710, 1724 

[6] Ofir, A. \etal (2013) - in prep.
}

\end{document}